# Off-axis gyration induces large-area circular motion of anisotropic microparticles in a dynamic magnetic trap

Yuan Liu,[1,†] Gungun Lin,[1,†] and Dayong Jin[1,2]

[1]Institute for Biomedical Materials and Devices, Faculty of Science, The University of Technology Sydney, Ultimo, New South Wales 2007, Australia
[2]UTS-SUStech Joint Research Centre for Biomedical Materials & Devices, Department of Biomedical Engineering, Southern University of Science and Technology, Shenzhen, 518055, China
†: these authors contributed equally to this work.
*Dr. Gungun Lin.
Email: gungun.lin@uts.edu.au

**Abstract:** Magnetic tweezers are crucial for single-molecule and atomic characterization, and biomedical isolation of microparticle carriers. The trapping component of magnetic tweezing can be relying on a magnetic potential well that can confine the relevant species to a localized region. Here, we report that magnetic microparticles with tailored anisotropy can transition from localized off-axis gyration to large-area locomotion in a rotating magnetic trap. The microparticles, consisting of assemblies of magnetic cores, are observed to either rotate about its structural geometric center or gyrate about one of the magnetic cores, the switching of which can be modulated by the external field. Raising the magnetic field strength above a threshold, the particles can go beyond the traditional synchronous-rotation and asynchronous-oscillation modes, and into a scenario of large-area circular motion. This results in peculiar retrograde locomotion related to the magnetization maxima of the microparticle. Our finding suggests the important role of the microparticle's magnetic morphology in the controlled transport of microparticles and developing smart micro-actuators and micro-robot devices.

Magnetic tweezing is a powerful technique to manipulate single or multiple nano/microparticles on demand[1,2], to concentrate particles from surrounding media[3,4], study the mechanical properties of biological samples[5,6] and characterize the physical properties of single molecules and atoms[7]. Microscale magnetic tweezing can be reliant on magnetic potential wells represented by local maxima of magnetic fields. The gradient magnetic fields can be produced by the pole shoes of electromagnets [8,9], permanent magnets[10] and any other ferromagnetic micro- and nanostructures[11,12].

The dynamic form of magnetic tweezing can leverage the spatial-temporal organization of the magnetic fields from more sophisticated electromagnets, to enable applications in magnetic resonance imaging[13] and magnetic particle imaging[14], and microrobotic control[15,16]. On the other hand, a simple form of dynamic magnetic tweezing can be constructed by a rotating permanent magnet. It has been known that such a rotating trap can drive magnetic microspheres[17,18] or dimer-type structures[19,20], into synchronous rotation at low frequencies and asynchronous oscillation at high frequencies when the particles cannot catch up with the pace of the rotating field. Despite the rotation transition, the range of the particles' motion is typically limited to a small area. Moreover, in conventional scenarios, the influence of the morphology of a particle on the tweezing effect is typically overlooked.

Here we report that the off-axis gyration of an anisotropic microparticle can lead to large-area circular locomotion using a pair of rotating permanent magnets (Figure 1a). The microparticles are fabricated with micro-sized magnetic core (s)-polymeric shell structures with sizes comparable to that of a magnetic potential well, using a two-step microfluidic emulsification approach described previously[19,21]. We find that the morphology of the magnetic particles can play a crucial role in modulating the magnetic trapping point on the microparticles. The rotation center of a microparticle is observed to switch between its structural geometric center and the magnetic core center at varied trapping potential. The mechanism enables the motion stance of a microparticle to switch among synchronous off-axis gyration/symmetric rotation, asynchronous rotation, and large-area retrograde motion (Figure 1b). The motion modes can be modulated by multiple parameters, such as magnetic field gradient, strength, frequency and the magnetic architecture of the microparticles.

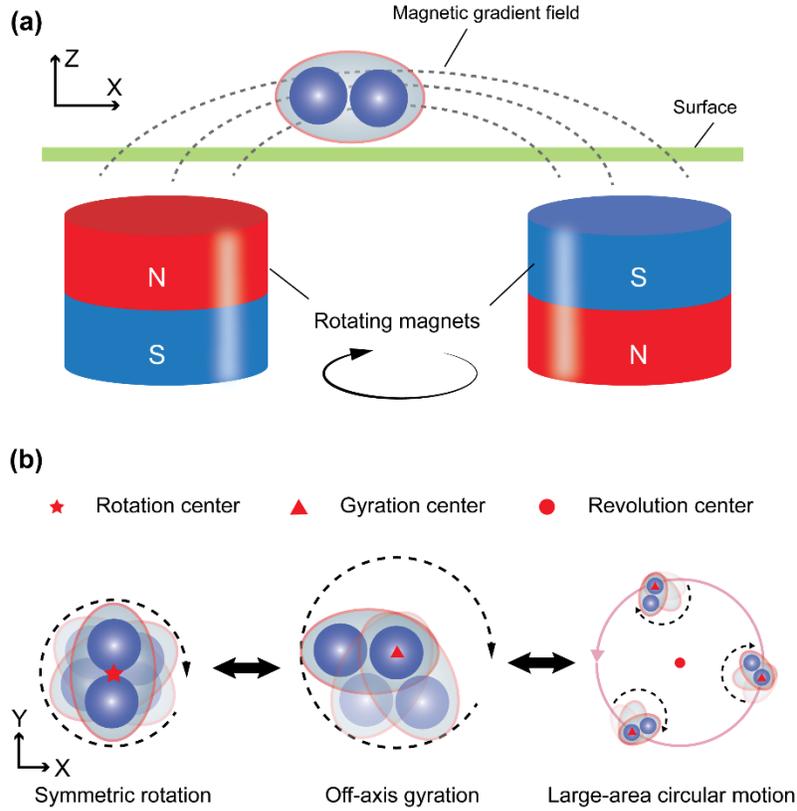

**Figure 1**. Tweezing anisotropic magnetic microparticles using rotating permanent magnets. (a) Schematic illustration of the experimental scenario of rotating a dual-core microparticle under a gradient magnetic field. The distance is the height of the microparticle to the magnets' plane. The speed of rotated magnets can be modulated from 0 to 1500 rpm. (b) Schematic illustration showing that a microparticle could undergo motion mode transition, including transition between symmetric rotation and off-axis gyration, and switching between off-axis gyration and large-area circular motion. The rotation centre of symmetric rotation locates at the geometry centre. The off-axis gyration has an off-axis gyrating radius. The large-area motion is hierarchical, consisting of gyration and circular revolution.

A dynamic magnetic trap can be constructed from a pair of rotating magnets embedded in a stirrer device. Using a 3D Hall sensor, the distribution of the magnetic fields (denoted by three major components) of the magnets were measured (data not shown). Computational simulation (COMSOL Multiphysics) of the magnets shows that the gradient field component, $\nabla B_x$, is represented by a dipolar-shaped pattern with local maximum and minimum (Figure 2a). This resulted in a magnetic potential well located at the center between the magnets. In a static state, the potential well, acting as a magnetic trap, attracts a magnetic particle towards its center. A comparable gradient field component along the z axis, i.e. $\nabla B_z$, can be found to superimpose on the x-component (Figure 2b). Different from $\nabla B_x$, the $\nabla B_z$ has a peak-like pattern and is symmetric with respect to the center of the magnets. The depth and width of the potential well can be adjusted by the distance of the microparticle's working surface from the magnets. While reducing distance, the potential well becomes deeper and narrower, indicating a stronger and more localized trapping force. This dependency suggests that when the size of a microparticle is comparable with that of the potential well, the particle can experience uneven magnetic forces. In this case, the particle's morphology may come into play and two force-equilibrium positions could be established. Specifically, a sizable particle could be trapped to the center of the potential well with its geometric center overlapping with the well center (here referring to Mode 1, as shown in Figure 2c). Alternatively, the particle could have its end trapped by the potential well, with the other end slightly off the surface, with a larger mismatch of its geometric center with the well center (here referring to Mode 2, as shown in Figure 2d).

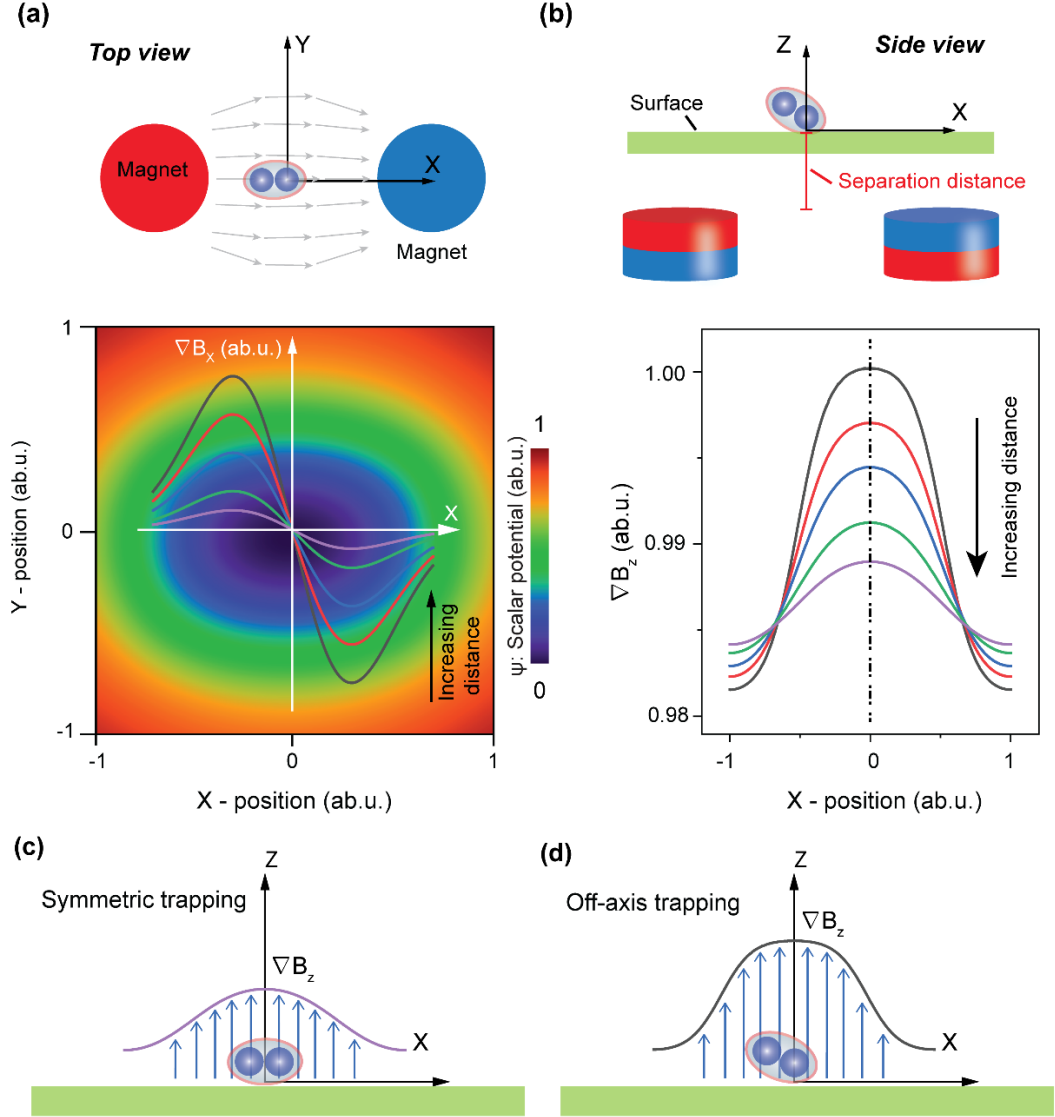

**Figure.2** Mechanism of transition between off-axis gyration and symmetric trapping under magnetic field-structure coupling. (a) X-axis component of magnetic field gradient. (Upper) Top view of the experimental configuration. (Bottom) X-axis component of the magnetic field gradient, $\nabla B_x$. Inserted curve shows the profile of the field gradient with different distance between the microparticle and magnets' plane. (B) Z-axis component of magnetic field gradient. (Upper) Side view of the experimental configuration. (Bottom) Z-axis component of the field gradient for different distance between the microparticle and the magnets' plane. Illustration of the equilibrium modes of symmetric trapping (C) and off-axis trapping (D). The former is featured with symmetric trapping where the particle is located with its geometric center overlapping with the center of the potential well. The latter is shown with the particle lying in an off-axis position with respect to the well center. The two modes correspond to two different potential well depths and widths.

To elaborate further the theoretical basis, the major magnetic tweezing force exerted on a purposely-designed dual-magnetic core microparticle can be decomposed into two components: gradient-field magnetic force and magnetic anisotropy force as approximated by a dipole-dipole coupling force.[20] For a larger separation distance from the magnets and a low magnetic field strength, the former component could become inappreciable in comparison with the latter. In this case, the dipolar-coupling induced torque can be balanced with the hydrodynamic torque of the fluid, $\tau_h = -\varepsilon \eta V \vec{\omega}_{ma}$, at a critical rotation frequency, as typically observed in previous studies. As the particle is mainly driven by the anisotropy induced torque, the particle will rotate around its geometric center at low field frequencies, as described in Figure 2c, Mode 1. Both permanent and induced magnetic moments may contribute to the torque[22,23]. We have demonstrated previously that the current system under investigation is majorly governed by the induced magnetic moments[19].

When the field strength is raised, the gradient-field force component, namely, the trapping force, could play a dominating role over the dipolar coupling force. A likely scenario is off-axis trapping of a microparticle as described by Figure 2d, Mode 2. This mode can be justified when a particle, in practice, may deviate from an ideally symmetric structure and isotropic composition. Positive $\nabla B_z$ tends to lift the particle at one end adjacent to the center, causing asymmetric trapping forces exerted on the microparticles. The difference in the trapping force is balanced by the friction force of the particle to establish an equilibrium state of magnetic trapping to one end of the particle.

The off-axis gyration, featured with a significant mismatch between the geometry center and rotation center, indicates that an increased centrifugal force can be associated with the gyration at high rotating speeds. In this case, at a certain threshold rotating speed, centrifugal force may overcome the gradient force and the friction force, causing the particle to slide off the center of the magnetic potential well. Consequently, an equilibrium state of circular motion can be achieved when the centrifugal force is balanced by the sum of magnetic trapping force and friction force exerted on the microparticle. Worth noting, the friction force is expressed as $F_s = \mu_s F_n$, where $\mu_s$ is the coefficient of kinetic friction and $F_n$ is the normal force directly perpendicular to their surface. During the sliding-off process, $F_n$ is a dynamic force decided by the magnetic lift force:

$$F_n = G - F_z - F_b \qquad [1]$$

where G is the gravity, $F_z$ is the lift force induced by the $B_z$, and $F_b$ is the buoyancy force.

The anisotropy force may drive the particle to gyrate while circling with a large radius. By balancing the period of the circulating motion with that of the rotating field, one may obtain the radius of large-area circular motion, which can be expressed as[24]:

$$R_{cm} \approx \frac{kl\omega_{mg}(1-p_{slip})}{2\pi\omega_{cm}} \qquad [2]$$

where $\omega_{mg}$, $\omega_{cm}$ is the angular speed of the external magnetic field and the circular motion, respectively; k the geometric factor, l is the length of the particle, $p_{slip}$ is the slipping probability defined by the error function of the ratio between the centrifugal force and magnetic force [24]. $p_{slip}$ is 0 at relatively slow rotation and approaches 1 at high rotational speed of magnetic field, for which the centrifugal force may far exceed the magnetic force. k is equal to ½ for symmetric rotation and may take on other values for off-axis gyration, depending on the position of the gyration centre. The above relationship suggests the feasibility to regulate the motion area by changing the rotation frequency.

To validate the above hypothesis, we designed two types of particles: one with two similar-sized cores with 300 μm in diameter and the other with two uneven cores with 200 and 400 μm in diameter. Under a low rotation frequency of 3 Hz, both particles are observed to exhibit synchronous rotation about their structural center at a low field strength of 10 mT rotating at 3Hz (Figure 3a). This is evidenced by the rotation radius equal to the sum of the radii of the cores. When the phase lag is smaller than $\pi/2$, rotation in this mode is synchronous with the external magnetic field below a critical frequency that scales with the magnetic field strength (Figure 3b). Above the critical frequency (phase lag > $\pi/2$), the particle may rotate backwards in a rotating cycle when it cannot catch up with the pace of the field (Figure 3c). This rotation mode switches to synchronous off-axis gyration, namely, rotating about one of its cores' center, at a larger field strength of 30 mT when the separation distance from the magnets is reduced (Figure 3d).

Remarkably, the particle with an uneven core size is shown to gyrate about the center of the larger core. This suggests that the magnetic potential well tends to grasp the core with stronger magnetization.

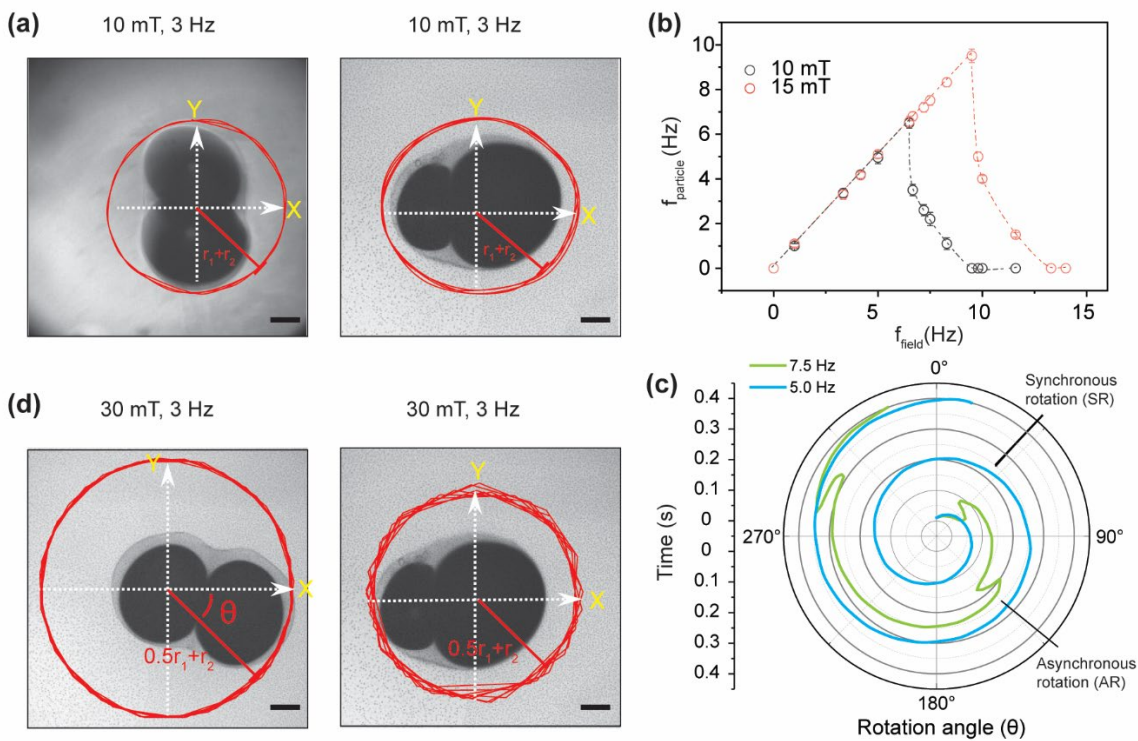

**Figure. 3** Experimental evidence of switching between symmetric gyration and off-axis gyration. (a) Tracking the rotation of two types of particles: one with two similar-sized cores with 300-μm diameter and the other with two uneven cores with 200 and 400 μm in diameter. Here $r_1$ and $r_2$ is the radius of each core, respectively. The red lines indicate the trajectories of symmetric rotation. The field strength of $B_x$ at the field centre is 10 mT and the field frequency is 3 Hz. (b) Dependence of the rotation frequency of the particle on the rotating field frequency at two magnetic field strengths: 10 mT and 15 mT. Error bar denotes the measurement accuracy. (c) Polar plot representation of the time-lapsed motion trajectory of a particle under different cycling frequencies of 5 Hz (green colored) and 7.5 Hz (blue colored). (d) Tracking of the off-axis gyration of the particles at a higher field strength of 30 mT. Scale bar in (a) and (d):100 μm.

Figure 4a shows that the microparticle can be driven to large-area circular motion at increased frequencies. At low frequencies such as below 7 Hz, it is shown that the radius of motion is constant, which indicates that the centrifugal force is not sufficient to overcome the magnetic trapping force. Large-area circular motion occurs at the rotating field frequency larger than 7 Hz, upon which the centrifugal force exceeds 0.3 nN. The motion is featured with a unique retrograding pace. The gyration center is observed to switch between two cores of the microparticles, resembling brachiation-like locomotion (Figure 4c). During the first half cycle (0°~180°), one core contacts the substrate with the other slightly lifted. The stance results in gyration around the center of the substrate-contacting core, and switches to the other during the second half gyration cycle. The clockwise gyration and gyration-centre switching result in the anticlockwise retrograde locomotion of the microparticle. A dynamic range of frequency regulation exists to alter the motion radius between 0 mm to 2.75 mm. Correspondingly, the centripetal forces of magnetic particles are derived to be 4.65 nN and 7.50 nN for 2.25-mm (4.30 s period) and 2.75-mm (3.70 s period) orbital radii, respectively. At even higher frequencies, an increased chance of slipping may occur, leading to a reduced locomotion speed for the dual-core particle (Dual-P). The three-core particle (Tri-P), however, is observed to locomote circularly at larger radii at all rotating field speeds. This could be ascribed to that the longer length of the particle can be featured with an increased length of pace during each gyration cycle at zero slipping probability, as described by Equation (3) (Figure 4b).

This is further confirmed by the fact that the locomotion speed of the Tri-P scales with that of the rotating field in the corresponding range up to 1400 rpm, as is in contrast with that of the Dual-P. Figure 4d displays the map of the locomotion stance based on the field frequency and field strength of $H_y$ at the centre of the magnets. Under low rotating speed and strength of the magnetic field, the microparticles exhibit the synchronous rotation (SR). We found that due to the difference in the magnetic trapping force, the synchronous rotation region may consist of two states: symmetric rotation and off-axis gyration. Once increasing the frequency, the off-axis gyration transits to the large-area circular motion (LAM) due to the large the centrifugal force, while the symmetric rotation transits to asynchronous rotation (SR) due to the large phase lag between the field's orientation and the particle's magnetization direction (Figure 4d).

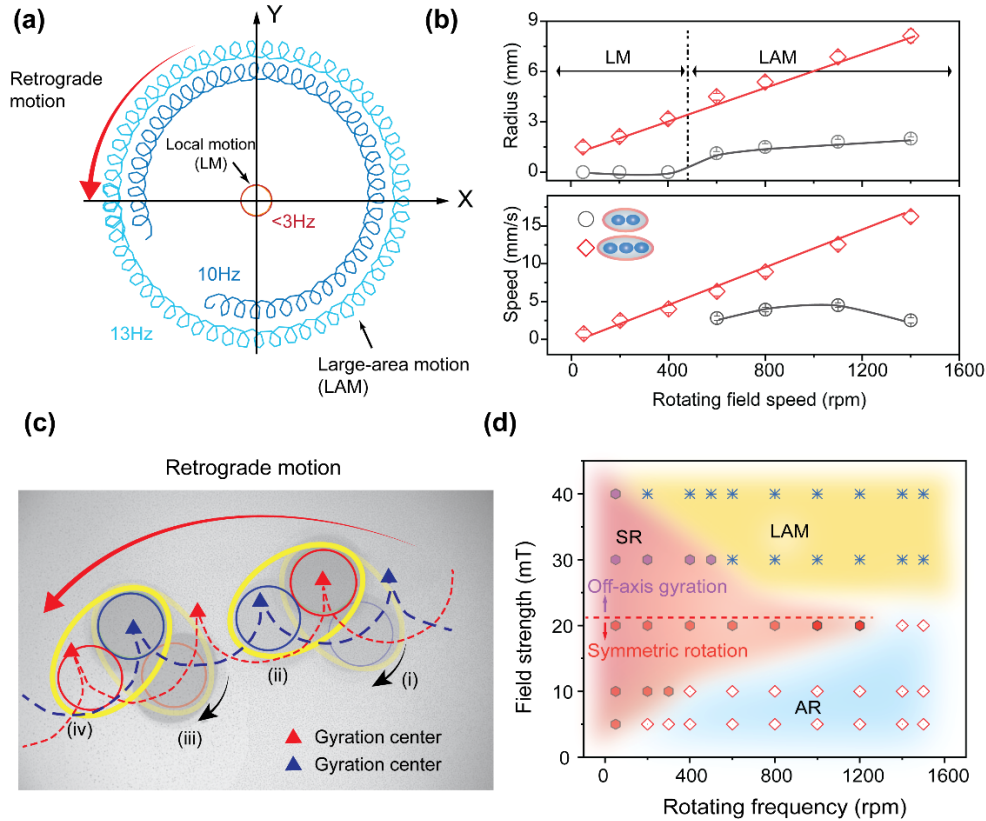

**Figure. 4** Characteristics of large-area circular motion. (a) Locomotion trajectories of local motion (LM) and large-area retrograde motion (LAM) of microparticles at different rotating frequencies. (b) Motion radius and speed changing with field speed. The results are obtained from the average values of 5 particles with identical core ratios. (c) Retrograde locomotion. The gyration center, marked by the triangular shape, switches in between two cores of the microparticles during the gyration cycle. (d) Mapping of motion modes of a particle with regard to the variation of field strength and field frequency. Three locomotion modes are included: synchronous (SR), asynchronous (AR) and large-area retrograde motion (LAM). The field strength of $B_x$ at the center is changed from 5 to 40 mT, and the field frequency ranges from 50 rpm to 1500 rpm.

In conclusion, we have presented the systematic study of the motion modes of anisotropic microparticles under a pair of rotating magnets. It has been found that the cooperative effect of the in-plane trapping force and out-of-plane lift force can unlock the off-axis gyration of anisotropic magnetic microparticles. This is manifested by the rotation center of a particle to switch from its geometric center and the center of one of its magnetic cores at a high trapping force. The off-axis gyration of the particles at high speeds can lead to sustainable and maneuverable large-area circular motion, which provides insight into the controlled transport of microparticles using magnetic traps.

G.L. acknowledges financial support from the National Health and Medical Research Council (GNT1160635). G.L. and D.J. thank the financial support of the ARC Industry Transformational Research Hub Scheme (grant IH150100028). Y. L. thanks the financial support of China Scholarship Council (201608140100).

DATA AVAILABILITY

Data available on request from the authors.